\documentclass{emulateapj}
\usepackage{epsfig}
\usepackage{graphicx}
\usepackage{color}
\usepackage{enumerate}

\newcommand{\msun}{\,\rm M_\odot}

\newcommand{\be}{\begin{equation}}
\newcommand{\ee}{\end{equation}}
\newcommand{\ba}{\begin{eqnarray}}
\newcommand{\ea}{\end{eqnarray}}
\newcommand{\f}{\frac} 
\newcommand{\rvir}{r_{\rm vir}}
\newcommand{\rtwo}{r_{\rm 200}}
\newcommand{\mvir}{M_{\rm halo}}

\newcommand{\Vmax}{V_{\rm max}}
\newcommand{\rVmax}{r_{\rm Vmax}}
\newcommand{\rt}{r_{\rm t}}

\newcommand{\kms}{\,{\rm km\,s^{-1}}}
\newcommand{\meanq}{\langle q \rangle}
\newcommand{\means}{\langle s \rangle}
\newcommand{\meanT}{\langle T \rangle}

\newcommand{\peakVmax}{V_{\rm max,p}}
\lefthead{}
\righthead{Shapes of dark matter subhalos}



\begin{document}
\submitted{}

\title{The shapes, orientation, and alignment of Galactic dark matter subhalos}

\author{Michael Kuhlen$^1$, J\"urg Diemand$^{2,4}$, \& Piero
Madau$^{2,3}$}

\affil{$^1$School of Natural Sciences, Institute for Advanced Study,
Einstein Lane, Princeton, NJ 08540 \\$^2$Department of Astronomy \&
Astrophysics, University of California, Santa Cruz, CA
95064\\$^3$European Southern Observatory, Karl-Schwarzschild-Str. 2,
D-85748 Garching, Germany\\$^4$Hubble Fellow}
%\email{mqk@sns.ias.edu, diemand@ucolick.org, pmadau@ucolick.org.}

\begin{abstract}
We present a study of the shapes, orientations, and alignments of
Galactic dark matter subhalos in the ``Via Lactea'' simulation of a
Milky Way-size $\Lambda$CDM host halo. Whereas isolated dark matter
halos tend to be prolate, subhalos are predominantly triaxial. Overall
subhalos are more spherical than the host halo, with minor to major
and intermediate to major axis ratios of 0.68 and 0.83,
respectively. Like isolated halos, subhalos tend to be less spherical
in their central regions. The principal axis ratios are independent of
subhalo mass, when the shapes are measured within a physical scale
like $\rVmax$, the radius of the peak of the circular velocity
curve. Subhalos tend to be slightly more spherical closer to the host
halo center. The spatial distribution of the subhalos traces the
prolate shape of the host halo when they are selected by the largest
$\Vmax$ they ever had, i.e. before they experienced strong tidal mass
loss. The subhalos' orientation is not random: the major axis tends to
align with the direction towards the host halo center. This alignment
disappears for halos beyond $3 \rtwo$ and is more pronounced when the
shapes are measured in the outer regions of the subhalos. The radial
alignment is preserved during a subhalo's orbit and they become
elongated during pericenter passage, indicating that the alignment is
likely caused by the host halo's tidal forces. These tidal
interactions with the host halo act to make subhalos rounder over
time.
\end{abstract}

 
\section{Introduction}

Cold dark matter (CDM) halos are not smooth. Early numerical
simulations of the formation of CDM halos lacked sufficient resolution
to detect much besides the gross features of the mass distribution
\citep{Davis1985, Frenk1985, Quinn1986, Efstathiou1988, Frenk1988}.
Rapid advances in computational power and in the efficiency of codes
have afforded a dramatic reduction in the particle masses and
gravitational softening lengths. Multi-mass techniques
\citep[e.g.][]{Katz1993, Bertschinger2001} allow even further
resolution increases in particular areas of interest. The resulting
high resolution maps of the matter distribution in CDM halos have
revealed an astonishing abundance of substructure \citep[][hereafter
paper I]{Klypin1999, Moore1999,Diemand2007a}. In ``Via Lactea'', the
most recent and highest resolution CDM simulation of a Galaxy scale
halo to date, the total number of identifiable subhalos has reached
$\sim 10,000$, which together account for $5.6\%$ of the total host
halo mass (paper I). Recent progress notwithstanding, the currently
achievable mass resolution is orders of magnitude above the true
cutoff in the CDM fluctuation power spectrum, and given the observed
scale-invariant nature of the subhalo mass function ($dN/d\ln M
\propto M^{-1}$), it is likely that the total substructure abundance
and mass fraction has not yet converged.

CDM halos are not round. Whereas analytical work often treats CDM
halos as spherically symmetric mass distributions, it has been known
for some time \citep{Barnes1987,Efstathiou1988, Frenk1988} that in
general CDM halo shapes show significant departures from
sphericality. There is now a large body of work concerning the shapes
of isolated CDM halos
\citep{Dubinsky1991,Katz1991,Warren1992,Dubinski1994,Jing1995,Tormen1997,Thomas1998,Bullock2002,Jing2002,Springel2004,Bailin2005,Kasun2005,Hopkins2005,Allgood2006,Bett2007,Hayashi2007,Maccio2007}
and widespread agreement on a number of findings: CDM halos tend to be
prolate, they are more spherical in their outer regions, more massive
halos tend to have smaller axis ratios, the moment of inertia is
aligned with the shape velocity anisotropy tensor, the smallest
principal axis tends to line up with the angular momentum vector,
etc. \citep[for a recent summary, see][]{Allgood2006}.

With Via Lactea's unprecedented resolution, we are now for the first
time able to extend this shape analysis to a well resolved subhalo
population. This is interesting from a theoretical point of view,
because of the important role that tidal interactions play in shaping
the properties of the subhalo population. The shapes of subhalos are
likely to be affected by tidal deformations, and a careful analysis of
the subhalo shapes might allow us to better understand the tidal
interactions.

Knowledge of subhalo shapes is important for several types of
observational studies as well. Weak lensing is becoming a very
valuable tool for probing cosmological parameters \citep{Brown2003,
Bacon2003, Hamana2003, Jarvis2003, Hoekstra2006, Massey2007} and
constraining density profiles of galaxy groups and clusters
\citep{Brainerd2004,Mandelbaum2006a}. Any alignment between intrinsic
galaxy ellipticity and with other galaxies or the local mass density
will introduce a bias into the lensing signal
\citep{Hirata2004,Bridle2007}. In fact, \citet{Lee2005} found evidence
for an intrinsic alignment, with subhalo minor axes preferentially
perpendicular to the host halo's major axis, in a numerical simulation
of four cluster scale halos, resolved with $\sim 2 \times 10^6$
particles each. They also developed an analytical model under the
assumption that this alignment was caused by the host halo's tidal
field, and showed that it reproduces the numerical findings
well. Observational evidence for subhalo alignment has be found in the
Sloan Digital Sky Survey (SDSS), for galaxies in clusters
\citep{Agustsson2006,Mandelbaum2006b} as well as in groups
\citep{Faltenbacher2007}. Based on N-body numerical experiments
\citet{Ciotti1994} concluded that the tidal field of a spherical
cluster could be responsible for the observed radial alignment of
cluster galaxy isophotes. A better understanding of typical subhalo
shapes and their alignment within the host halo might allow a
statistical correction of the resulting bias.

Another observational arena dependent on subhalo shapes are stellar
kinematical studies of Local Group dwarf galaxies, which are being
used to constrain the masses of their DM host halos
\citep{Wilkinson2004,Lokas2005,Munoz2005,Walker2006,Gilmore2007,Strigari2007}.
In almost all cases the analysis is performed assuming spherical
symmetry and often also a constant velocity anisotropy. More
sophisticated analyses of the dwarf galaxy stellar motions will
benefit from firm theoretical expectations of the intrinsic DM subhalo
shapes and should result in more realistic models of the underlying DM
mass distribution.

This paper is organized as follows. In Section~\ref{sec:method} we
briefly outline the technique we employed to determine the shape of
subhalos. We present the shape parameters of the Via Lactea host halo
in Section~\ref{sec:hosthalo_shape}, and move on in
Section~\ref{sec:subhalo_shape} to discuss the dependence of the
subhalo's shape parameters on the radius at which they are measured,
on their distance from the center of the host, and on the subhalo's
mass. In Section~\ref{sec:subhalo_alignment} we present results
concerning the spatial distribution of the subhalos within the host
halo and the alignment of their ellipsoids towards the host halo
center. In Section~\ref{sec:evolution} we consider the redshift
evolution of the shapes of a sample of strongly tidally affected
subhalos. Section~\ref{sec:conclusion} contains a discussion and
summary of our results.

\section{Simulation and Shape Finding Method} \label{sec:method}

The main sample of subhalos analyzed in this work stems from the $z=0$
output of the ``Via Lactea'' simulation (paper I). This simulation
follows the dark matter substructure of a Milky-Way-scale halo
($M_{200}=1.77 \times 10^{12} \msun$) with 234 million particles. With
a particle mass of $\simeq 20,000 \msun$, the simulation resolves
around 10,000 subhalos within $\rtwo=388$ kpc. The global $z=0$
properties of the host halo and the substructure population was
presented in paper I and the joint temporal evolution of host halo and
substructure properties, with an emphasis on tidal interactions, was
discussed in \citet[][hereafter paper II]{Diemand2007b}. Here we focus
on the shapes of the matter distribution in Via Lactea's host halo and
subhalo population.

For the determination of (sub)halo shapes we follow the iterative
technique outlined in \citet{Allgood2006}. This method is based on
diagonalizing the weighted moment of inertia tensor
\begin{equation}
\tilde{I}_{ij} = \sum_n \f{x_{i,n} x_{j,n}}{r_n^2},
\label{eq:MoI}
\end{equation}
where
\begin{equation}
r_n = \sqrt{ x_n^2 + (y_n/q)^2 + (z_n/s)^2 }
\label{eq:rellip}
\end{equation}
is the ellipsoidal distance in the eigenvector coordinate system
between the (sub)halo's center and the n$^{\rm th}$ particle, and
$q=b/a$ and $s=c/a$ are the intermediate-to-major and minor-to-major
axis ratios, respectively ($a \geq b \geq c$). Initially
$\tilde{I}_{ij}$ is calculated for all particles within a spherical
window of radius $r_0$. In subsequent iterations we fix $a=r_0$ and
include only particles with $r_n < r_0$. Iteration continues until $q$
and $s$ change by less than $10^{-3}$. The degree of triaxiality of a
halo is quantified by the triaxiality parameter introduced by
\citet{Franx1991}
\begin{equation}
T = \f{1-q^2}{1-s^2}.
\end{equation}
A halo is said to be oblate for $T<1/3$, triaxial for $1/3<T<2/3$, and
prolate for $T>2/3$.

For comparison with observational data it is often more desirable to
constrain the shape of the potential than the shape of the density
distribution. The potential shape has the additional advantage that it
is less sensitive to local density variations (from substructure for
example) and is typically smooth and well approximated by concentric
ellipsoids \citep{Springel2004, Hayashi2007}. Instead of measuring the
potential shape directly, or by fitting ellipses to the intersections
of isopotential surfaces with three orthogonal planes \citep[as
advocated by][]{Springel2004}, we diagonalize the unweighted kinetic
energy tensor
\begin{equation}
K_{ij} = \f{1}{2} \sum_n v_{i,n} v_{j,n}, 
\end{equation}
where the $v_{i,n}$ are measured in the restframe of the subhalo under
consideration. $K_{ij}$ is related to the potential energy tensor
$W_{ij} = \sum x_i d\Phi/dx_j$ through the tensor virial theorem
\begin{equation}
\f{1}{2} \f{d^2I_{ij}}{dt^2} = 2 K_{ij} + W_{ij}.
\end{equation}
The internal structure of the Via Lactea host halo remains practically
unchanged after $z=1.7$ (paper II) and we expect $d^2I_{ij}/dt^2=0$ at
$z=0$. In that case the eigenvectors of $K_{ij}$ should reflect the
principal axes of the potential ellipsoid. Note that the assumption of
a constant $I_{ij}$ is likely to fail for subhalos that are being
tidally stripped, and in this case the principal axes of $K_{ij}$ will
not necessarily reflect the shape of the potential. Instead of the
iterative procedure applied for $I_{ij}$, we simply diagonalize
$K_{ij}$ for all particles within the moment of inertia ellipsoid at a
given radius. In the following we will refer to shapes determined by
diagonalization of $I_{ij}$ and $K_{ij}$ as the \textit{mass} and
\textit{potential} shapes, respectively.

\section{Host Halo Shape} \label{sec:hosthalo_shape}

\begin{figure*}
\begin{center}
\includegraphics[height=0.6\textheight]{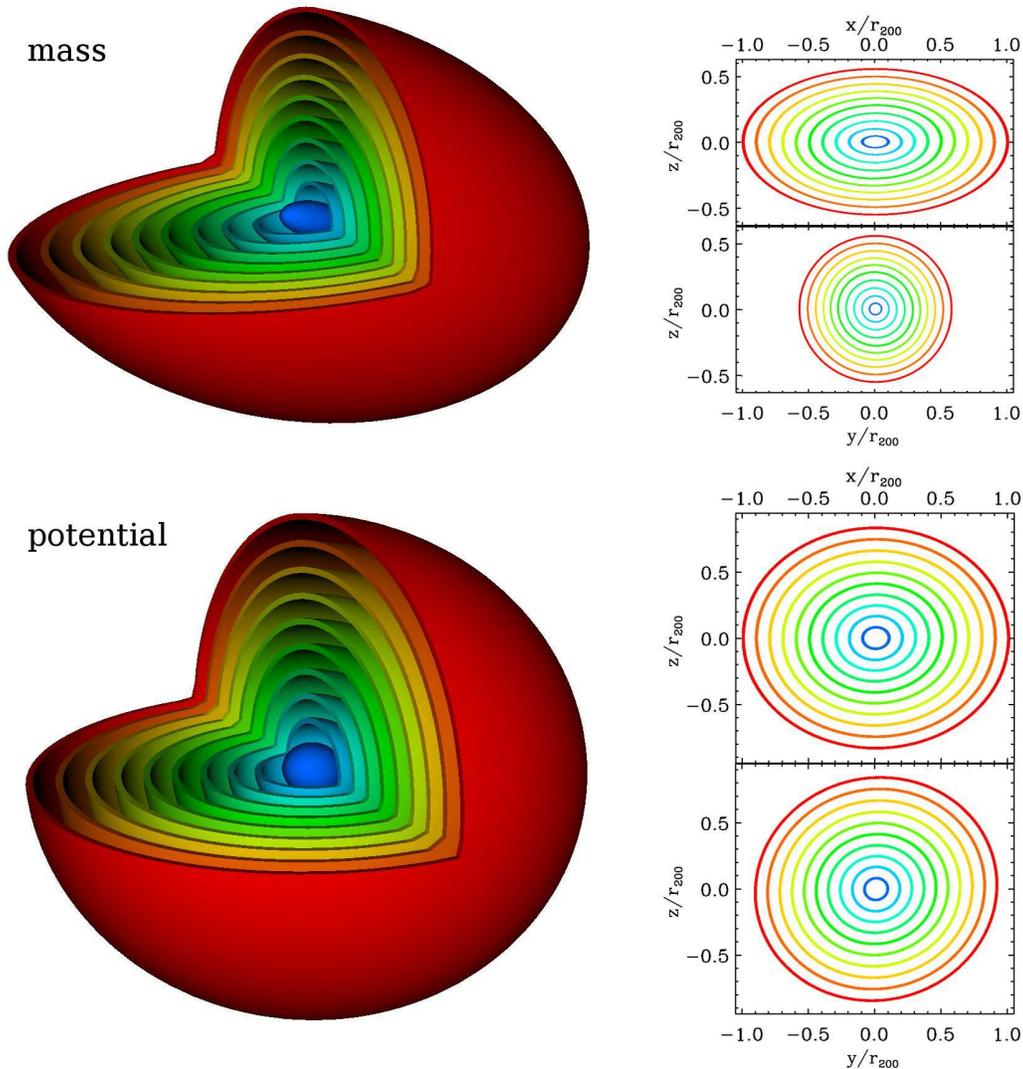}
\caption{The shape of the Via Lactea halo as a function of radius,
derived from diagonalizing the moment of inertia tensor $I_{ij}$
(\textit{top}), and the velocity anisotropy tensor $K_{ij}$ as a proxy
for the shape of the potential (\textit{bottom}). The color indicates
radius. The major principal axes at different radii are aligned to
within $1^\circ$, and the mass and potential shapes are aligned to
within $5^\circ$.}
\label{fig:ellipsoids}
\end{center}
\end{figure*}

\begin{figure}
\includegraphics[width=\columnwidth]{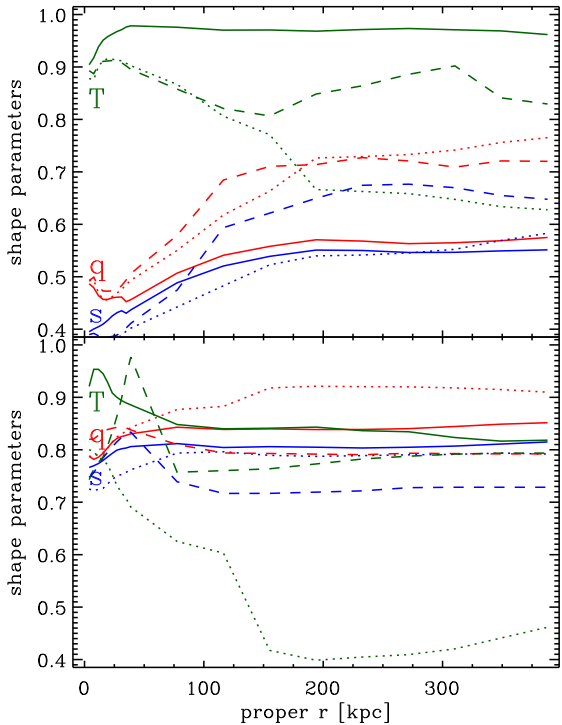}
\caption{Host halo shape parameters for the mass distribution
(\textit{top panel}) and the potential (\textit{bottom panel}) as a
function of proper distance from the halo center at $z=0$
(\textit{solid lines}), $z=0.5$ (\textit{dashed lines}), and $z=1$
(\textit{dotted lines}). Intermediate to major axis ratio $q$
(\textit{red}), minor to major axis ratio $s$ (\textit{blue}), and
triaxiality parameter $T$ (\textit{green}).}
\label{fig:hosthalo_shape}
\end{figure}

At the present epoch the Via Lactea host halo is prolate. We have
determined the principal components of $I_{ij}$ and $K_{ij}$ as a
function of major axis radius. The resulting shape ellipsoids are
depicted in Fig.~\ref{fig:ellipsoids}. Neither the shape nor the
orientation of the ellipsoids vary much out to $\rtwo$. As a function
of radius the major axis eigenvectors for both mass and potential
shape are aligned to within $1^\circ$. As expected the potential is
significantly rounder than the mass distribution, and the major axes
of the two are aligned to within 5$^\circ$ at all radii. In
Fig.~\ref{fig:hosthalo_shape} we plot mass and potential shape
parameters $q$ and $s$ and triaxiality parameter $T$ at $z=$0, 0.5,
and 1 as a function of radius. Recall (from paper II, Figs.~3 and 4)
that the spherically averaged mass distribution of the Via Lactea host
halo remains remarkably constant \textit{in proper coordinates} after
the last major merger at $z=1.7$ until today. We now see, however,
that the host halo does undergo some adjustments in its shape.

At $z=0$ the mass distribution becomes slightly less spherical towards
the center with axis ratios dropping from $\sim 0.55$ at 200 kpc $<r<
\rtwo$ to $\sim 0.45$ at 20 kpc, whereas the potential shape axis
ratios remains constant at around 0.8. The triaxiality parameter
remains firmly in the prolate regime ($>2/3$) for both mass and
potential throughout the entire halo. The potential shape becomes
slightly more prolate in the inner regions. At earlier times, however,
we detect some significant changes in the axis ratios, especially in
the outer regions. At $z=0.5$ both $q$ and $s$ are larger by about
0.1-0.15 beyond 100 kpc, and at $z=1$ $q$ is significantly larger than
$s$, resulting in a triaxial, as opposed to prolate, outer region
beyond $\sim200$ kpc. The potential shape exhibits variations of
comparable magnitude. Note that the host halo accretes some fairly
massive subhalos ($M_{\rm sub}/M_{\rm host} \sim 0.1$) between $z=1$
and $z=0.5$. Dynamical friction causes these subhalos to spiral in to
the center over a few orbits, and they lose most of their mass
($>99\%$) in this process. The associated redistribution of material
probably contributes to the observed shape adjustments.

The shape of the Via Lactea host halo is consistent with previous
studies of the shapes of dark matter halos, which have generally found
them to be mostly prolate \citep[e.g][ and references
therein]{Allgood2006}. Observational studies of the shape of the Milky
Way galaxy, however, have been much less conclusive. Disk warping
\citep{Olling2000} and some models of the Sagittarius tidal stream
\citep{Ibata2001, Majewski2003, MartinezDelgado2004} suggest that the
Milky Way halo is close to spherical ($q \approx s \gtrsim 0.8$) and
oblate, whereas some studies of the leading arm of the Sagittarius
stream favor a prolate shape with $s=0.6$ \citep{Helmi2004, Law2005}.

The collisionless nature of our Via Lactea simulation prohibits a
direct comparison between our findings and the observational
constraints on the shape of the Milky Way host halo. Previous
hydrodynamical numerical studies of galaxy formation have found that
the cooling of baryons leads to significantly rounder halos
\citep{Kazantzidis2004, Springel2004, Bailin2005}, especially in the
central regions where s can increase by as much as 0.2 to 0.3. At the
moment it is unclear how much of this effect is due to a potential
overcooling problem in these simulations.

\section{Subhalo Shapes} \label{sec:subhalo_shape}

We now turn to the analysis of the shapes of Via Lactea's
subhalos. Unlike the host halo, which is resolved by over ten million
particles even within $0.1 \rtwo$, the subhalo shape determination is
limited by numerical resolution. We restrict our analysis to halos
containing at least 5,000 particles within their tidal radius,
$\rt$. As in paper I, we define $\rt$ as the point where the subhalo's
spherically averaged density profile drops to twice the local
underlying matter density. Via Lactea contains 97 such halos within
$\rtwo$, 309 within $3 \rtwo$, and 829 in total.

One complication arises from the difficulty in distinguishing between
actual subhalo particles and those belonging to the underlying host
halo. Our subhalo finder (6DFOF, cf. paper I) assigns all particles
within the tidal radius to the subhalo, without removing unbound
particles.  Some contribution to $I_{ij}$ and $K_{ij}$ would thus come
from background particles. This contribution is generally negligible
for $I_{ij}$, since the background particles are more or less
uniformly distributed throughout the subhalo. The host halo particles,
however, significantly distort $K_{ij}$, because their velocities are
typically much larger and more anisotropic than the subhalo particle
velocities, especially for subhalos close to the host halo center. For
this reason we have cleaned the subhalos from background host halo
particles by comparing the members of each subhalo to those of its
progenitor in the penultimate simulation output at $z=0.005$ ($\sim
68.5$ Myr before $z=0$), and retained only those particles that also
appear within $\rt$ at the earlier time.  This effectively removes all
host halo particles from the subhalo and allows an accurate
determination of $K_{ij}$. We have confirmed that the shape of
$I_{ij}$ is not affected by this cleaning procedure.

The subhalo tidal radius shrinks as subhalos pass through the inner
regions of the host halo due to the increasing background
density. However, as we showed in paper II, not all
particles beyond this reduced $\rt$ are actually stripped from the
subhalo, and $\rt$ re-expands as the subhalo begins its climb out of
the host halo potential. Measuring the shape for all particles within
the tidal radius thus probes more central regions for subhalos closer
to the halo center than for subhalos in the outer regions. To avoid
this bias we focus on the subhalo shapes for all particles within the
radius of the peak of the circular velocity curve, $\rVmax$. This
choice comes at the cost of a reduced number of particles, but
$\rVmax$ does not temporarily decrease at pericenter passage.

\subsection{Shape vs. Radius}
\label{sec:shape_vs_radius}

\begin{figure}
\includegraphics[width=\columnwidth]{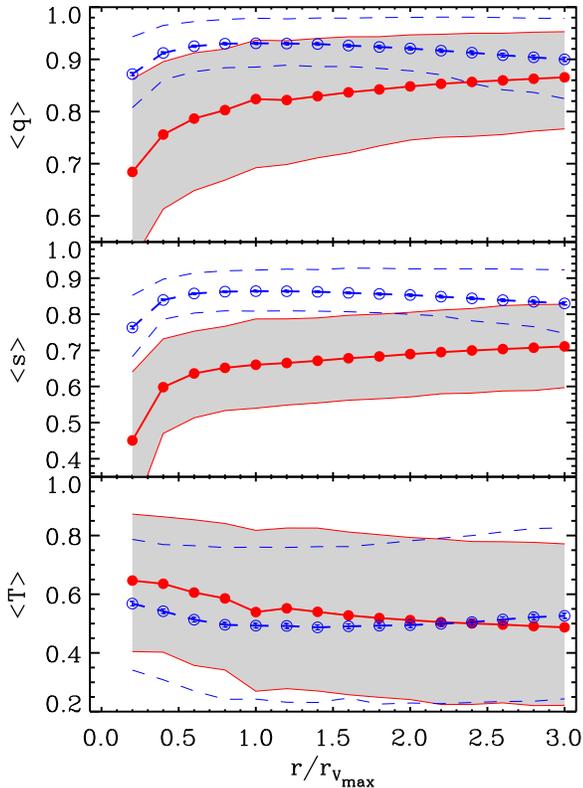}
\caption{Mean (sub)halo shape parameters vs. the major axis radius of
the ellipsoidal window used in the shape determination. We plot the
shape of the moment of inertia tensor $I_{ij}$ (\textit{solid lines,
filled symbols}) and the shape of the kinetic energy tensor $K_{ij}$
(\textit{dashed lines, open symbols}).
%Error bars indicate the uncertainty in the mean and the
The 68\% scatter around the mean is indicated by the shaded region for
$I_{ij}$ and by the thin dashed lines for $K_{ij}$. \textit{Top
panel}: intermediate to major axis ratio, $q=b/a$. \textit{Middle
panel}: minor to major axis ratio $s=c/a$. \textit{Bottom panel}:
triaxiality parameter, $T=(1-q^2)/(1-s^2)$.}
\label{fig:shape_vs_radius}
\end{figure}

We first discuss the dependence of subhalo shape on the radius at
which they are measured, see Fig.~\ref{fig:shape_vs_radius}. For this
analysis we included the complete sample of 829 halos with more than
5000 particles within their tidal radius. Most of these halos
currently lie outside of $\rtwo$, but a significant fraction (78\%
within $2\rtwo$) have passed through the host halo at some earlier
time (paper II) and have experienced tidal interactions. Similarly to
the Via Lactea host halo, we find that the subhalos' mass
distributions become slightly less spherical in the inner regions;
$\meanq$ decreases from about 0.87 at $3\rVmax$ to 0.68 at
$0.2\rVmax$, the innermost point at which we determine subhalo
shapes. Similarly, $\means$ decreases from 0.72 to about 0.45. The
68\% scatter is about 0.2 for both $q$ and $s$. As expected the
potential shape shows even less dependence on radius and is almost
spherical, with $\meanq \approx 0.9$ and $\means \approx 0.85$. The
potential axis parameters also have a smaller 68\% scatter of about
$0.1$.

Unlike isolated dark matter halos, which tend to be prolate, we find
that subhalos are generally triaxial. $\meanT$ shows a slight
decreasing trend with radius, but remains in the triaxial regime for
both mass and potential shape throughout the range of radii that we
probed. Note that the 68\% scatter around $\meanT$ extends into both
the prolate and oblate regime. At $\rVmax$ about 25\% of all
(sub)halos are prolate and an equal number oblate.

\subsection{Shape vs. Distance}

\begin{figure}
\includegraphics[width=\columnwidth]{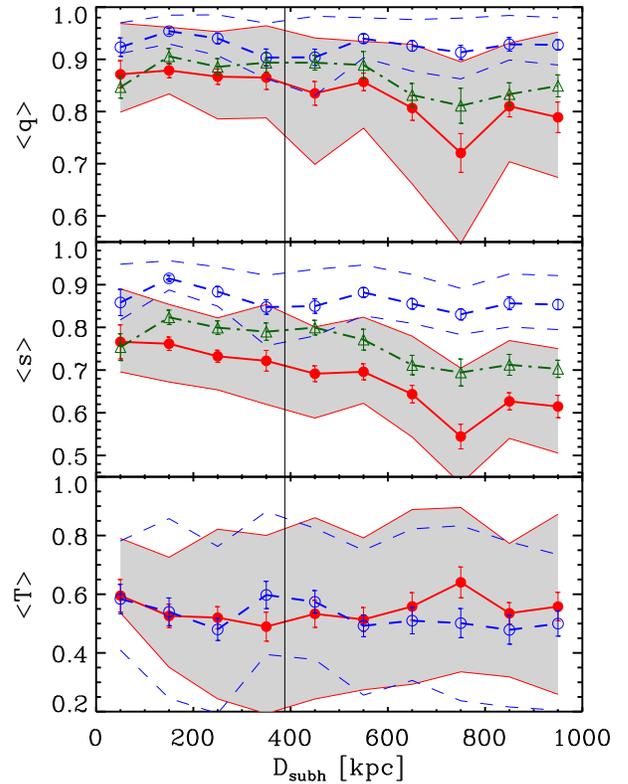}
\caption{Mean (sub)halo shape parameters measured within $\rVmax$ as
a function of distance from the halo center. In addition to the mass
shape (\textit{solid lines, filled circles}) and the potential shape
(\textit{dashed lines, open circles}), in the top two panels we also
plot the shape parameters measured within $\rt$
(\textit{dash-dotted lines, open triangles}). The thin solid line
indicates $\rtwo=388$ kpc.  }
\label{fig:shape_vs_distance}
\end{figure}

Next we take a look at the dependence of subhalo shapes on the
distance from the halo center. For this purpose we use the shapes
determined from all particles within
$\rVmax$. Fig.~\ref{fig:shape_vs_distance} shows that the shape of the
mass distribution is close to independent of distance, with a weak but
significant trend towards slightly larger axis ratios closer to the
host halo center. Within $\rtwo$ the mean axis ratios are
$\meanq=0.87$ and $\means=0.74$, and outside of $\rtwo$ they are
$\meanq=0.81$ and $\means=0.64$. The potential shape is independent of
distance, with $\meanq=0.93$, $\means=0.86$. The ellipsoids remains
predominantly triaxial over the whole range of distance, for both mass
and potential.

One might have expected a stronger distance dependence, since the
subhalo shapes are affected by tidal interactions, which are stronger
at the center. This is likely a consequence of measuring the shapes
within $\rVmax$, which is deep inside the halo mass distribution, and
not as strongly affected by tides. Measuring the shapes within the
tidal radius should give a more pronounced effect, and indeed, when we
look at the trend vs. distance for shapes measured within $\rt$
(dashed lines and open triangles in Fig.~\ref{fig:shape_vs_distance})
we find that for the innermost bin the mean axis ratios decrease. At
least part of this trend, however, may also be explained by the
temporary decrease of $\rt$ as the subhalo passes through pericenter
and the shape-radius dependence discussed in
Section~\ref{sec:shape_vs_radius}.

\subsection{Shape vs. $V_{\rm max}$}

\begin{figure}
\includegraphics[width=\columnwidth]{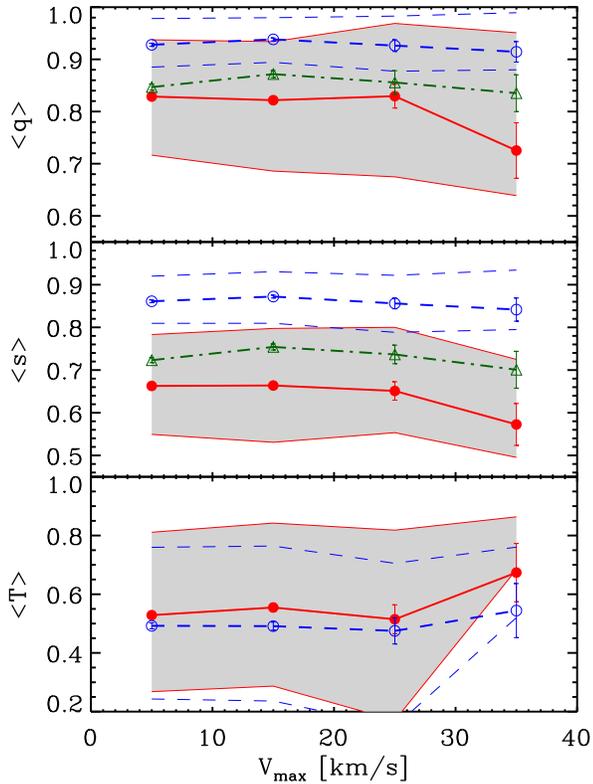}
\caption{Subhalo shape parameters as a function of the subhalos peak
circular velocity, a proxy for mass. Panels and symbols as in
Fig.~\ref{fig:shape_vs_distance}.}
\label{fig:shape_vs_vcmax}
\end{figure}

Lastly, we consider the shape parameters as a function of the
subhalo's $\Vmax$, a proxy for mass,
Fig.~\ref{fig:shape_vs_vcmax}. Again we use all 829 (sub)halos in the
simulation. \citep{Allgood2006} found that the axis ratios of isolated
field halos measured at $0.3 \rvir$ decreased with increasing halo
mass, i.e. less massive halos are more spherical. In contrast, we find
no such trend for subhalos. With the exception of the highest mass
bin, both $q$ and $s$ appear to be independent of $\Vmax$.

The highest $\Vmax$ bin (centered on $35 \kms$, corresponding to a
tidal mass of $\sim10^{10} \msun$) has $\means=0.57 \pm 0.05$, a bit
less than would be expected from an extrapolation of the
\citet{Allgood2006} $\means-\mvir$ relation which gives $\means
\approx 0.69 \pm 0.04$ at $10^{10} \msun$ for $\sigma_8=0.74$. All but
one of the six halos in this bin lie within $2\rtwo$ and can thus be
considered subhalos. In \citet{Allgood2006} shapes were measured at
$0.3 \rvir$, whereas our subhalo shapes are measured at $\rVmax$. We
find it difficult to assign a ``virial'' radius to subhalos (paper
II), and instead use the tidal radius $\rt$ as the outer ``edge'' of
the subhalo. The median ratio of $\rVmax$ to $\rt$ in this bin is
0.16, and for a $10^{10} \msun$ field halo with a concentration of 15
$\rVmax / \rvir = 2.163/c = 0.14$, so it is likely that we are simply
measuring the shapes farther in, where axis ratios tend to be smaller
(see Fig.~\ref{fig:shape_vs_radius}). For comparison, $\means=0.70$
when measured at $\rt$.

At any rate, our mean subhalo axis ratios become independent of mass
at lower $\Vmax$. It is possible that tidal interactions cause this
flattening of the subhalo shape-mass relationship. Another possibility
is that halo shapes (for both subhalos and field halos) are in fact
intrinsically independent of mass when measured at a fixed physical
scale, such as $\rVmax$. In this case the mass dependence found at
$0.3 \rvir$ would in effect just be combination of radius dependence
and the halo mass -- concentration relation: $0.3 \rvir$ is smaller
than $\rVmax$ in massive, low concentration halos. This means that
relative to $\rVmax$, the shapes of more massive halos are probed at
smaller radii, where axis ratios tend to be smaller. Note that in this
picture also the redshift dependence observed by \citet{Allgood2006}
might in part be a consequence of a window ($0.3 \rvir$) which becomes
larger with time due to its co-moving definition. Further
investigations will be necessary to fully address this issue.

\section{Subhalo Distribution and Shape Alignment} \label{sec:subhalo_alignment}

Having discussed the dependence of subhalo shapes on their properties,
we now consider the spatial distribution and alignment within the host
halo.

\subsection{Spatial Distribution}

The spatial distribution of subhalos depends sensitively on the sample
selection criterion. In paper II we showed that the $z=0$ distribution
of Via Lactea subhalos is anti-biased compared to the mass
distribution of the host halo. For subhalos selected to have a $z=0$
mass greater than $4 \times 10^6 \msun$, the ratio of subhalo number
density to host halo mass density is simply proportional to radius,
whereas for subhalos selected to have a $z=0$ $\Vmax$ greater than $5
\kms$, this ratio scales as the enclosed mass. Previous studies
have found that when subhalos are selected by their mass or $\Vmax$
at the time of accretion, this bias is substantially reduced, and the
resulting subhalo number density profile more closely traces the
underlying mass distribution \citep{Nagai2005, Faltenbacher2006}.

Here we extend this analysis to a full 3D spatial distribution of the
subhalos. In addition to the two samples considered in paper II, we
include a range of samples selected by the highest $\Vmax$ a subhalo
reaches throughout its lifetime, a quantity we refer to as
'$\peakVmax$'. Three of these samples are defined by lower limits,
$\peakVmax >$ 5, 10, and 15 $\kms$, and three more by including
the 100, 500, 1000 subhalos with the largest $\peakVmax$.  This type
of selection is designed to remove the bias introduced by tidal
interactions, since the selection is performed on subhalo properties
prior to their interaction with the host. For each sample of subhalos
we diagonalize a weighted ``moment of inertia'' tensor
(Eq.\ref{eq:MoI}) constructed from the $z=0$ positions of the
subhalos, without regard for their masses. We then calculate the axis
ratios of the resulting ellipsoids and the angles between their
principal axes and the principal axes of the host halo's mass
distribution measured at $\rtwo$. 

\begin{table}
\caption{The 3D spatial distribution of subhalos}
\label{tab:subhalo_ellipsoids}
\begin{tabular}{ccccccc}
\hline
\hline
 & $N_h$ & $q$ & $s$ & $\theta_a$ & $\theta_b$ & $\theta_c$ \\
\hline
                    host halo &  --  & 0.58 & 0.55 &   -- &   -- &   -- \\
 $M > 4 \times 10^6 \msun$ & 5404 & 0.66 & 0.54 & 10.1 & 36.2 & 34.7 \\
        $\Vmax > 5 \kms$ & 1571 & 0.68 & 0.58 & 13.5 & 20.1 & 15.9 \\
    $ \peakVmax > 5 \kms$ & 3824 & 0.64 & 0.59 & 8.36 & 15.1 & 12.8 \\
   $ \peakVmax > 10 \kms$ &  689 & 0.65 & 0.55 & 5.90 & 7.70 & 5.09 \\
   $ \peakVmax > 15 \kms$ &  224 & 0.64 & 0.52 & 10.6 & 3.13 & 10.9 \\
 top 100 $\peakVmax$ &  100 & 0.61 & 0.41 & 10.6 & 9.55 & 12.9 \\
 top 500 $\peakVmax$ &  500 & 0.67 & 0.54 & 4.47 & 4.42 & 2.91 \\
top 1000 $\peakVmax$ & 1000 & 0.64 & 0.57 & 4.10 & 9.02 & 9.63 \\
\end{tabular}
\tablecomments{The shapes of the subhalo spatial distribution for
different subhalo samples (described in the text). $N_h$ is the number
of subhalos in the sample, $q$ and $s$ the axis ratios, and
$\theta_{a,b,c}$ the angle between the major, intermediate, and minor
principal axis of the subhalo and the host halo ellipsoid.}
\end{table}

The spatial distribution of all subhalo samples is triaxial, with $s$
comparable to the underlying host halo density distribution and
slightly larger $q$. Not all of the ellipsoids, however, are aligned
with the host halo. The major axis of the ellipsoid defined by all
subhalos with $M_t > 4 \times 10^6 \msun$ ($\Vmax > 5 \kms$) is tilted
by $10.1^\circ$ ($13.5^\circ$) with respect to the host halo's major
axis. In general, the subhalo samples selected by $\peakVmax$ are more
closely aligned with the host halo. The best alignment is found for
the sample consisting of the 500 subhalos with the largest
$\peakVmax$, whose principal axes are tilted by less than $5^\circ$
from the host halo's. The results for all samples are summarized in
Table~\ref{tab:subhalo_ellipsoids}. Note that \citet{Zentner2005},
\citet[][2007]{Libeskind2005}, and \citet{Kang2007} also find that
the substructure distribution is well aligned with the host halo
orientation.

\begin{figure}
\includegraphics[width=0.9\columnwidth]{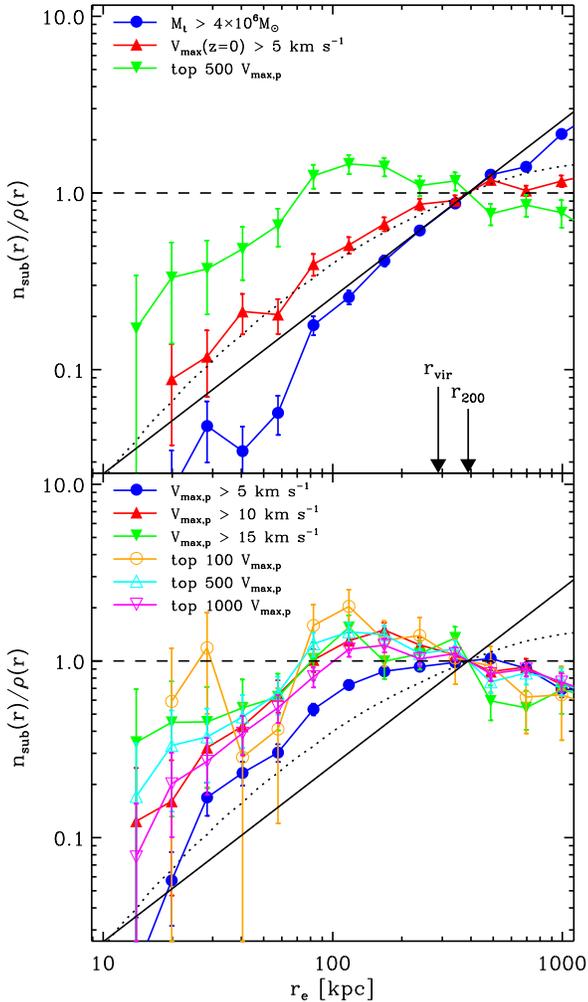}

\caption{The ratio of the subhalo number density $n_{\rm sub}(r_e)$ to
the host halo mass density $\rho(r_e)$ as a function of ellipsoidal
radius, $r_e = \sqrt{(x')^2 + (y'/q)^2 + (z'/s)^2}$. The ratio has
been normalized to unity at $r_e = \rtwo$. The solid line indicates
$n_{\rm sub}/\rho \propto r$ and the dotted line $n_{\rm sub}/\rho
\propto M(<r)$.}
\label{fig:subhalo_spatial_bias}
\end{figure}

The radial dependence of the subhalo number density is presented in
Fig.~\ref{fig:subhalo_spatial_bias}, where we plot the ratio of the
subhalo number density to the host halo's mass density as a function
of $r_e$, the ellipsoidal radius (\ref{eq:rellip}). The ratio has been
normalized to unity at $r_e = \rtwo$, in order to highlight the radial
dependence in the interior of the host halo. The top panel clearly
shows the radial dependence published in paper II hold also for
ellipsoidal binning: subhalo samples selected according to their
present mass or $\Vmax$ are anti-biased with respect to the host halo
mass distribution, with the former scaling as $n_{\rm sub}/\rho
\propto r$ down to $\sim 60$ kpc and the latter as $n_{\rm sub}/\rho
\propto M(<r)$ for the entire range of radii probed. The bottom panel
shows that a selection based on $\peakVmax$ removes a lot of this
anti-bias. All six samples have a radial number density dependence
closer to the host halo mass distribution than the samples selected
according to $z=0$ properties. The more restrictive the selection
criterion is, the larger $n_{\rm sub}/\rho$ becomes. For some of the
selections the ratio even exceeds unity from 80 kpc to $\rtwo$,
indicating that these samples are spatially biased with respect to the
host halo, i.e. their abundances falls off with radius faster than the
underlying density distribution. This bias is probably a consequence
of dynamical friction. The subhalos with the largest $\peakVmax$ were
massive enough to experience some degree of dynamical friction and
spiraled in towards the center. Along the way they lost mass, reducing
the dynamical friction and preventing them from completely merging
with the host. This process would preferentially remove such subhalos
from the outer regions.

The ``top 500 $\peakVmax$'' sample comes closest to tracing $\rho(r)$
(both in radial dependence and in the orientation of the shape
ellipsoid), and we have plotted it in the top panel for direct
comparison with the samples discussed in paper II. The decrease in
$n_{\rm sub}/\rho$ below $r_e=70$ kpc is probably at least partially
due to numerical resolution, since the decrease is smaller for larger,
and therefore better resolved subhalos.

\subsection{Radial Alignment}

\begin{figure}
\includegraphics[width=\columnwidth]{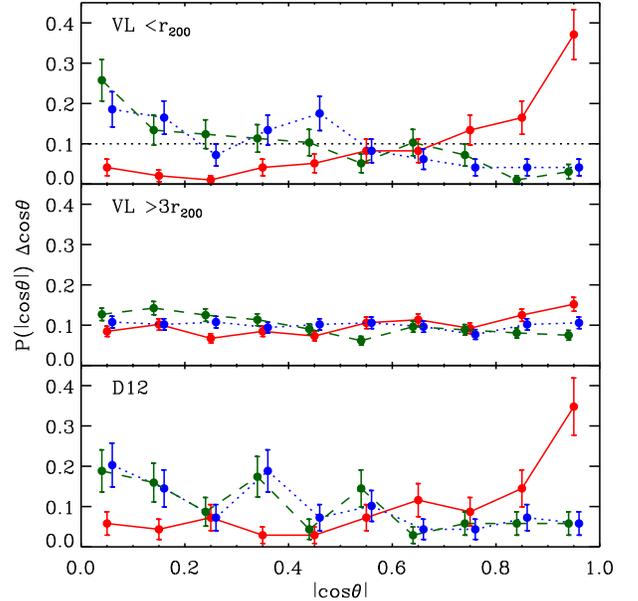}
\caption{The distribution of the cosine of the angle between the
direction towards the halo center and the subhalos' major
(\textit{solid lines}), intermediate (\textit{dotted lines}), and
minor (\textit{dashed lines}) axis, for all subhalos within
$\rtwo$. The data points indicate the fraction of subhalos within each
bin of width 0.1. \textit{Top panel}: 97 Via Lactea subhalos within
$\rtwo$.  \textit{Middle panel}: 520 Via Lactea subhalos outside of
$\rtwo$. \textit{Bottom panel}: 69 ``D12'' cluster subhalos within
$\rvir$. The intermediate and minor axis data points have been offset
by $\pm 0.01$ in the horizontal direction for clarity.}
\label{fig:alignment}
\end{figure}

Via Lactea's very high numerical resolution allows us to investigate
the orientation of dark matter subhalos with respect to the host halo
center. We find strong evidence for preferential radial alignment of
the subhalo triaxial mass distribution.

In the top panel of Fig.~\ref{fig:alignment} we plot the
distribution of $|\cos{\theta}|$, the absolute value of the cosine of
the angle between each of the principal axes of the subhalo's triaxial
ellipsoid and the radius vector from the host halo center, for the 97
Via Lactea subhalos within $\rtwo$. If the subhalo ellipsoids were
randomly distributed with respect to the halo center, this
distribution would be flat. Instead we see that the major axis
distribution is strongly peaked towards large values of
$|\cos{\theta}|$, indicating that the major axis preferentially points
towards the halo center. Correspondingly, the intermediate and minor
axes are biased towards low values of $|\cos{\theta}|$.

In the middle panel of Fig.~\ref{fig:alignment} we plot the same
distribution for all halos outside of $3\rtwo$. The distribution is
flat, showing no evidence for alignment of any of the principal axes.
The alignment effect appears to only be present for subhalos
physically associated with the Via Lactea host halo, and this is one
piece of evidence for a tidal origin of this radial alignment.

For comparison we have plotted (lower panel Fig.~\ref{fig:alignment})
the same distribution for the subhalos of a galaxy cluster scale dark
matter halo, the 14 million particle ``D12'' cluster discussed in
\citet{Diemand2004}. The same subhalo shape alignment is present here
too. \citet{Ragone2007} and Faltenbacher et al. (2007, in preparation)
also observe such an alignment in a much larger sample of lower
resolution group and cluster scale dark matter halos and their subhalo
population.

\begin{figure}
\includegraphics[width=\columnwidth]{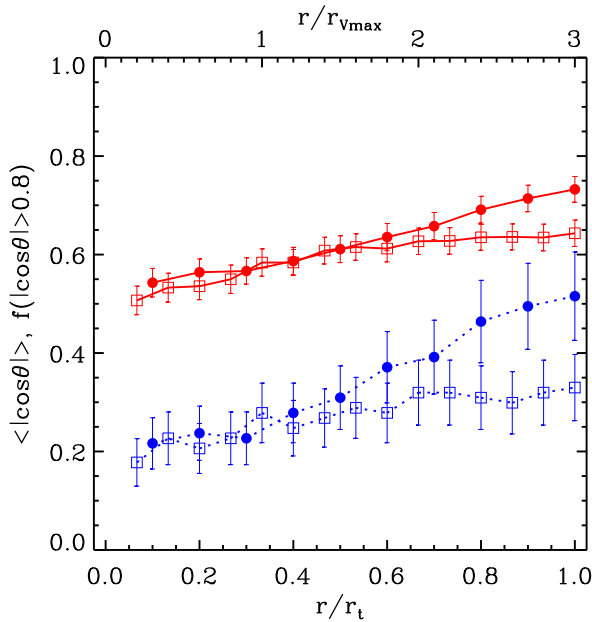}
\caption{The dependence of the subhalo alignment towards the host halo
center as function of subhalo radius at which the shape is
measured. Only subhalos within $\rtwo$ have been
included. \textit{Solid line}: the mean of $|\cos{\theta}|$ and its
uncertainty. \textit{Dotted line}: the fraction of subhalos with
$|\cos{\theta}|>0.8$ and its Poisson errors. Radii as a fraction of
$\rt$ (\textit{filled circles}, lower axis) and as a fraction of
$\rVmax$ (\textit{open squares}, upper axis).}
\label{fig:costheta_vs_radius}
\end{figure}

If tidal interactions are indeed responsible for the radial alignment,
then it may be expected that the alignment would be more pronounced in
the outer regions of the subhalo, where tidal effects are
strongest. Indeed we observe just such a trend when we look at the
mean of $|\cos{\theta}|$ and $f(|\cos{\theta}|>0.8)$, the fraction of
subhalos with $|\cos{\theta}|>0.8$, versus subhalo radius, see
Fig.~\ref{fig:costheta_vs_radius}. For this analysis we restrict
ourselves to subhalos within $\rtwo$, and consider radii as a fraction
of both $\rt$ and $\rVmax$. When measured at $\rt$, about 55\% of all
subhalos have a major axis that is aligned to within $35^\circ$ of the
direction towards the halo center and $\langle |\cos{\theta}| \rangle
= 0.75$. Both $\langle |\cos{\theta}| \rangle$ and
$f(|\cos{\theta}|>0.8)$ decrease monotonically towards smaller
$r/\rt$. The mean tidal radius for the 97 subhalos in this sample is
$\langle \rt \rangle = 10.6$ kpc, and $\langle \rVmax \rangle = $2.31
kpc. Not surprisingly, the alignment signal is less pronounced when
shapes are measured at $\rVmax$. However, even in this case the
alignment is significant: $\langle |\cos{\theta}| \rangle = 0.58$ with
an uncertainty ($=\sqrt{{\rm var}(|\cos{\theta}|)/N_{\rm sub}}$) of
0.03, which corresponds to a $\sim 2.5 \sigma$ significance.

Note that while \citet{Lee2005} finds that the subhalo minor axes are
preferentially perpendicular to the host halo's major axis in a study
of cluster-scale subhalos, and Faltenbacher et al. (2007, in
preparation) find a similar but weaker alignment signal in group-scale
subhalos, we find no evidence for such an alignment in the Via Lactea
subhalo population. The minor axes of our subhalo population shows no
correlation with any of the host halo's principal axes.

\section{Redshift Evolution} \label{sec:evolution}

The fact that the radial subhalo alignment is stronger for subhalos
closer to the host halo center (see Fig.~\ref{fig:alignment}) and that
the alignment signal is more pronounced when the shapes are measured
in the outer regions of the subhalo (see
Fig.~\ref{fig:costheta_vs_radius}) is suggestive of a a tidal origin
of the alignment. In this section we present additional support for
this hypothesis by looking at the temporal evolution of the shapes and
orientations of a small sample of subhalos, chosen according to the
following criteria:
\begin{enumerate}[i)]
\item the subhalos must lie within $\rtwo$ at $z=0$,
\item they must have undergone at least three pericenter passages
since $z=1.7$,
\item they must have experienced significant tidal mass loss, $\Delta
M/M = 1.0 - M_t(z=0)/M_t(z=1.7) > 0.4$, and
\item they must contain more than 4000 particles at $z=0$ ($M_t(z=0) >
8 \times 10^7 \msun$).
\end{enumerate}
These constraints result in a sample of 19 well resolved subhalos that
have experienced significant tidal interactions with the host halo.

\begin{figure}
\includegraphics[width=\columnwidth]{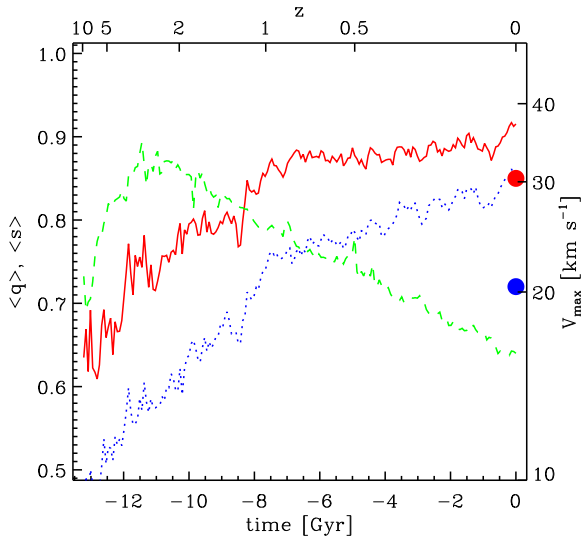}
\caption{$\meanq$ (\textit{solid line, red}), $\means$ (\textit{dotted
line, blue}), and $\langle \Vmax \rangle$ (\textit{dashed line,
green}) for the 19 subhalo sample (see text for a discussion of their
selection) as a function of time. The two filled circles at $z=0$
indicate $\meanq$ and $\means$ for all halos beyond $3\rtwo$, which
have been less affected by tides from the host halo. The shapes are
determined within $\rt$.}
\label{fig:mean_qs_vs_time}
\end{figure}

\begin{figure*}
\includegraphics[width=\textwidth]{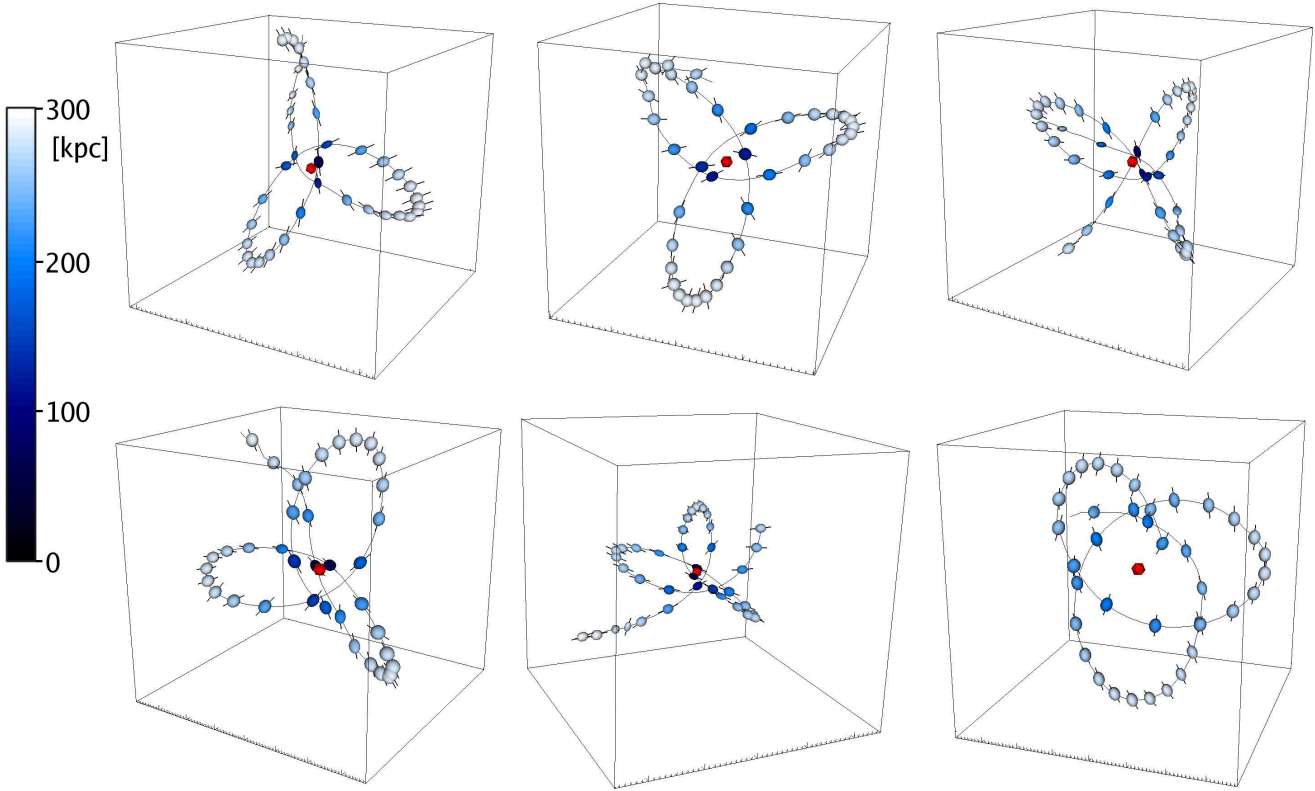}
\caption{The proper space orbits and shapes of six subhalos (see text
for a discussion of their selection) as function of time. Colors
indicate the distance in proper kpc from the host halo center, which
is indicated by the red icosahedron. The symbols depict ellipsoids of
a constant size (major principal axis of 10 kpc) whose axis ratios and
orientations are determined from all particles within the subhalo's
tidal radius. Ellipsoids are plotted for outputs between $z=2$ and
$z=0$ with a stride of 4 outputs, corresponding to a time interval of
274 Myr. The major principal axes are indicated by short solid lines.}
\label{fig:subhalo_orbits}
\end{figure*}

\begin{figure*}
\includegraphics[width=\textwidth]{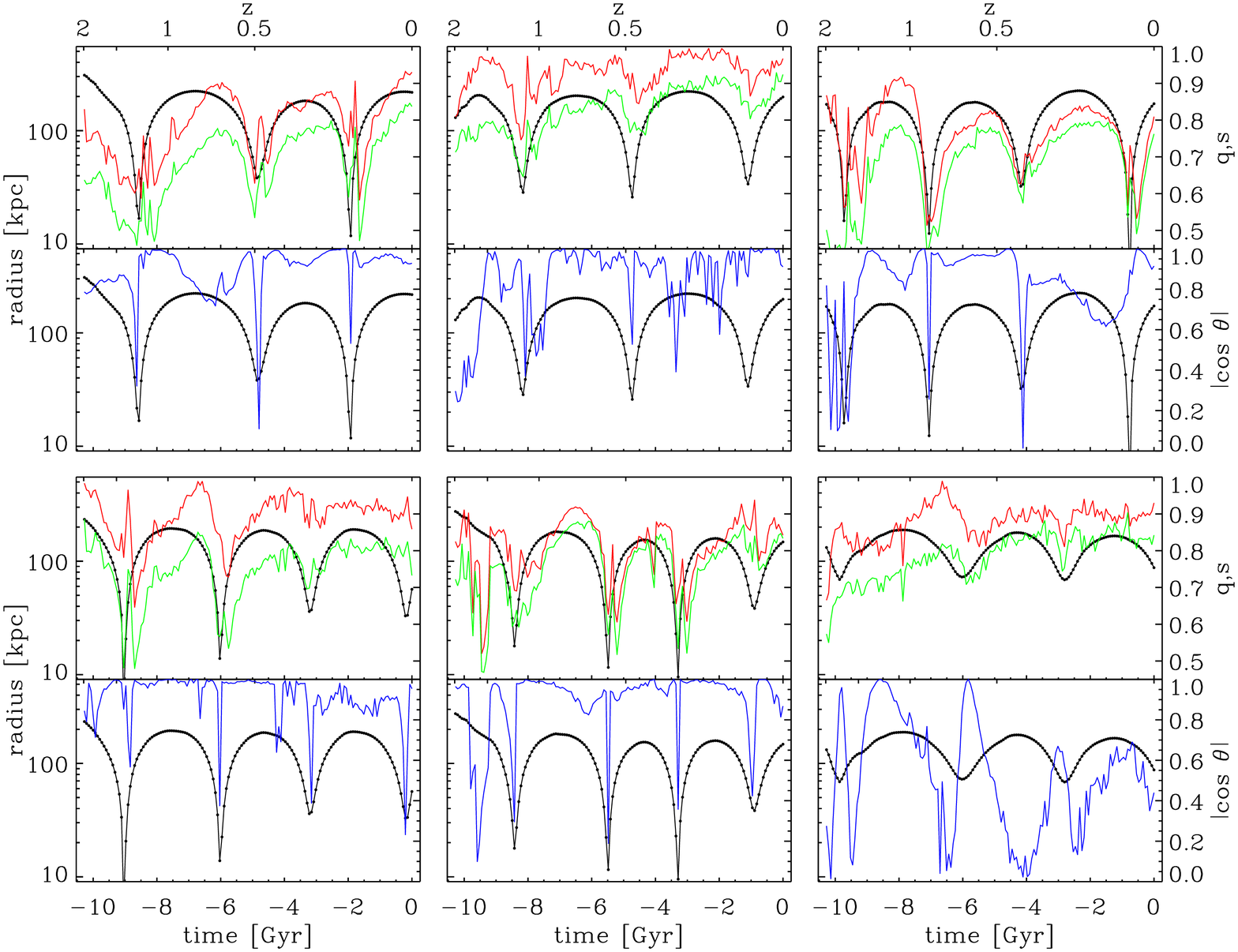}
\caption{The shape parameters (measured within $\rt$) and alignment of
six subhalos (see text for a discussion of their selection) as a
function of time. In each plot the subhalo's proper distance from the
host halo is shown in the black solid line. For each subhalo the top
panel also shows $q$ (\textit{upper line, red}) and $s$ (\textit{lower
line, green}) and in the bottom panel $|\cos{\theta}|$, the angle
between the subhalo's major axis and the direction towards the halo
center.}
\label{fig:shape_evolution}
\end{figure*}

In Fig.~\ref{fig:mean_qs_vs_time} we show the temporal evolution of
$\meanq$, $\means$, and $\langle \Vmax \rangle$ for this
subsample. The $\langle \Vmax \rangle$ curve shows that these subhalos
experience most of their mass growth early on and then continually
lose mass due to tidal interactions with the host halo (see paper II
for a more extensive discussion). Here we show that this tidal
interaction also leads to subhalos becoming rounder with time. At
formation they have $\meanq \approx 0.65$ and $\means \approx 0.5$,
but these grow as they are captured and begin to feel the host halo's
tidal field. After $z=1$ $\meanq$ remains stable at about $0.9$, but
$\means$ is still increasing slightly, reaching $\sim 0.85$ at
$z=0$. For comparison we also show the $z=0$ mean axis ratios of the
520 halos outside of $3\rtwo$. Owing to their large distance from the
host halo, these halos have experienced weaker tidal interactions. As
expected these halos are less spherical than the tidally stripped
sample: $\meanq=0.85$ and $\means=0.72$. This further supports the
notion that tidal interactions tend to make subhalos rounder.

In the following we further restrict our sample, and look in more
detail at the orbits of individual subhalos and the time dependence of
their shapes and alignments. For this purpose we hand selected five
illustrative examples of subhalos with small pericenters ranging from
9.6 to 32 kpc. For comparison we also included one subhalo (Id
\#21500) with a higher angular momentum orbit and a pericenter of 69
kpc. The properties of the selected subhalos are summarized in
Table~\ref{tab:evolution_sample}.

In Fig.~\ref{fig:subhalo_orbits} we present a three-dimensional
visualization of the orbits for the six subhalos in this
sample. Starting with the $z=0$ output and for every fourth output
thereafter (corresponding to a time interval of 274 Myr) up to $z=2$,
we have plotted at the subhalo's center of mass location an ellipsoid
whose orientation and axis ratios are determined from all particles
within the subhalo's tidal radius. The major axis of each ellipsoid is
indicated by a short solid line. For clarity we used a fixed major
axis length of 10 kpc, although the tidal radii vary along the
orbit. The orbits are shown in proper coordinates in the rest frame of
the host halo and the sides of the box range from -300 to +300 kpc.

The most striking feature of these plots is that the radial alignment
of the subhalo is preserved throughout the majority of its orbit. The
subhalos' ellipsoids perform a near perfect figure rotation such that
the major axis continually points close to the host halo's
center. This figure rotation, however, is not seen in subhalo \#21500,
the one with the higher angular momentum orbit. Its orientation is
almost independent of its orbital position.

\begin{table}
\caption{Subhalo properties of the shape evolution subsample}
\label{tab:evolution_sample}
\begin{tabular}{ccccccc}
\hline
\hline
        & \multicolumn{2}{c}{$z=0$} & \multicolumn{2}{c}{$z=1.7$} & & \\
Subhalo & $M_{t}$ & $\Vmax$ & $M_{t}$ & $\Vmax$ & $\Delta M/M$ \\
Id & ($\msun$) & ($\kms$) & ($M_\odot$) & ($\kms$) & \\
\hline
04242 & $2.4 \times 10^8$ & 15.4 & $3.7 \times 10^9$ & 28.2 & 0.93 \\
10876 & $1.6 \times 10^8$ & 14.6 & $4.6 \times 10^8$ & 18.8 & 0.66 \\
13351 & $4.4 \times 10^8$ & 18.8 & $2.6 \times 10^9$ & 66.9 & 0.84 \\
13467 & $1.2 \times 10^8$ & 14.3 & $2.8 \times 10^9$ & 28.6 & 0.96 \\
18412 & $1.3 \times 10^8$ & 12.5 & $4.0 \times 10^9$ & 32.5 & 0.97 \\
21500 & $8.4 \times 10^7$ & 11.9 & $1.5 \times 10^8$ & 13.4 & 0.45 \\
\end{tabular}
%\tablecomments{}
\end{table}

We have quantified these trends in Fig.~\ref{fig:shape_evolution},
where we plot each subhalo's radius, its axis ratios $q$ and $s$, and
its $|\cos{\theta}|$ as a function of time. The evolution of $q$ and
$s$ very closely tracks the orbit of the subhalo. Every time a subhalo
passes through pericenter its axis ratios decrease. Note that part of
this decrease can be attributed to the fact that the tidal radii
shrink during pericenter passage, and so we are effectively measuring
the subhalo shapes further in, where they are intrinsically less
spherical (see Section~\ref{sec:shape_vs_radius}). However, $\rt$
drops below $\rVmax$ only for one or two outputs close to pericenter,
and for most of the orbit remains in the regime where the axis ratios
are almost independent of radius. Thus we conclude that the temporary
decrease in $\rt$ at pericenter is not sufficient to explain the full
extent of the correlation between axis ratios and orbital
position. Furthermore we see from the plot of $|\cos{\theta}|$ that
the subhalos are indeed pointing close to the host halo center for the
majority of their orbits. During the fast pericenter passages
$|\cos{\theta}|$ drops significantly, but remains close to unity almost
everywhere else. There are a few exceptions, where $|\cos{\theta}|$
drops below unity even away from pericenter, for example at $z=0.7$
for \#04242 and at $z=0.4$ for \#10876 and \#13467, and these are most
likely caused by close passages to other massive subhalos.

Together these trends provide strong evidence for a tidal origin of
the radial alignment of subhalos. As they orbit the host halo's center
of mass, tidal forces continually distort the subhalos' mass
distribution, stretching them along the radial direction and
compressing them in the perpendicular directions. This tidal
distortion is stronger the closer the subhalo gets to the host halo
center, but the pericenter passage occurs so quickly that the subhalo
does not have enough time to adjust its orientation to point towards
the center.

Our subsample, of course, was chosen to have experienced strong tidal
interactions by requiring $\Delta M/M > 0.4$ and at least 3 pericenter
passages. However, as we showed in paper II, more than half of all
subhalos lose more than 50\% of their mass from $z=1$ to $z=0$, and so
tidally caused radial alignment is expected for most subhalos. As
mentioned above, subhalo \#21500 is one example of a subhalo that does
not appear to experience much radial alignment. We found that of the
19 subhalos for which we performed a time dependent analysis, only
three subhalos exhibit a similar lack of alignment-orbit correlation.

%\section{Discussion}

\section{Conclusion} \label{sec:conclusion}

The main conclusions of this paper can be summarized in the following points.
\begin{itemize}

\item The shape of the Via Lactea host halo is prolate, and slightly
less spherical (lower axis ratios) in the central regions. The shape
of the potential ($q \approx s \approx 0.8$) is significantly rounder
than the mass distribution ($q \approx s \approx 0.4-0.55$).

\item Whereas isolated halos tend to be prolate, we find that subhalos
are predominantly triaxial. Overall subhalos are more spherical than
the host halo. The mass shape has $\meanq=0.83$ and $\means=0.68$,
averaged over all subhalos within $3\rtwo$. The potential shape for
these subhalos is very close to spherical, with $\meanq=0.93$ and
$\means=0.86$. Like isolated halos, subhalos tend to be slightly less
spherical in the central regions.

\item We find a weak trend towards larger axis ratios for subhalos
closer to the host halo center. Within $\rtwo$ $\meanq=0.87$ and
$\means=0.74$, compared with $\meanq=0.82$ and $\means=0.65$ for all
subhalos outside of $\rtwo$.

\item For subhalos with $\Vmax < 30 \kms$ the axis ratios are
independent of $\Vmax$. At higher $\Vmax$ they are slightly lower.

\item The spatial distribution of subhalos matches the prolate shape
of the host halo when subhalos are selected by $\peakVmax$, the
largest $\Vmax$ they ever had during their lifetime. This type of
selection also results in an ellipsoidal radius dependence of the
subhalo abundance that more closely follows the mass distribution of
the host halo than the anti-biased distributions from selections based
on $z=0$ subhalo properties like $M_t$ and $\Vmax$.

\item The orientation of subhalo shape ellipsoids is not random. The
major principal axis of the subhalo mass distribution tends to align
with the direction towards the halo center. This alignment disappears
for subhalos beyond $\sim 3\rtwo$ and is more pronounced when the
shape is measured in the outer regions of the subhalo.

\item Tidal interactions with the host halo tend to make the subhalos
rounder over time.

\item For the majority of subhalos whose temporal evolution we studied
here in detail, the radial alignment is preserved during the subhalo's
orbit, and the axis ratios decrease during pericenter passage. We
conclude that the radial subhalo alignment is likely caused by tidal
interactions with the host halo.

\end{itemize}

\acknowledgments Support for this work was provided by NASA grants
NAG5-11513 and NNG04GK85G. J.D. acknowledges support from NASA through
Hubble Fellowship grant HST-HF-01194.01. The Via Lactea simulation was
performed on NASA's Project Columbia supercomputer system. The
visualizations in Figs.~\ref{fig:ellipsoids} and
\ref{fig:subhalo_orbits} were created with VisIt
(\texttt{http://www.llnl.gov/visit/}), a visualization tool developed
by the DOE's Advanced Simulation and Computing Initiative.

{}

\end{document}